\documentclass[10pt]{article}
\usepackage{graphicx}
\usepackage{amsmath}
\usepackage{amssymb}
\usepackage{caption2}
\begin{document}
\begin{center}
{\large {\bf \sc{  Analysis of the axialvector $B_c$-like tetraquark states with the QCD sum rules }}} \\[2mm]
Zhi-Gang  Wang \footnote{E-mail: zgwang@aliyun.com.  }     \\
 Department of Physics, North China Electric Power University, Baoding 071003, P. R. China
\end{center}

\begin{abstract}
In this article,  we construct the diquark-antidiquark type current operators to study the axialvector  $B_c$-like tetraquark states  with
the QCD sum rules. In calculations, we take the energy scale formula  as a powerful constraint to choose the ideal energy scales of the QCD spectral densities and add detailed discussions to illustrate  why we take the energy scale formula to improve the QCD sum rules for the doubly heavy tetraquark states. The predicted masses $M_{Z_{\bar{b}c}(1^{+-})}=7.30\pm0.08\,\rm{GeV}$ and $M_{Z_{\bar{b}c}(1^{++})}=7.31\pm0.08\,\rm{GeV}$ lie between the conventional $2\rm{P}$ and $3\rm{P}$ $B_c$ states with $J^P=1^+$ from the potential quark models, which is a typical feature of the diquark-antidiquark type $B_c$-like tetraquark states.
 \end{abstract}



\section{Introduction}

The mass spectrum of the $B_c$ mesons have been studied extensively in several theoretical  approaches,  such as
   the relativized (or relativistic) quark model with a phenomenological potential \cite{GI,EFG,GJ,ZVR}, the  nonrelativistic quark model with  a phenomenological potential \cite{GKLT,EQ,LuQF-potential},  the semi-relativistic quark model using the shifted large-$N$ expansion \cite{IS2004},  the lattice QCD \cite{Latt}, the Dyson-Schwinger equation and Bethe-Salpeter equation \cite{Changlei}, etc. Experimentally,
   only the ground state $B_c$ meson  with $J^{PC}=0^{-+}$ is listed in The Review of Particle Physics \cite{PDG}. Recently,
the CMS collaboration observed two excited $B_c$ states, which are consistent with the $B_c(\rm 2S)$ and $B_c^*(\rm 2S)$ respectively, in the $B_c^+\pi^+\pi^-$ invariant mass spectrum  in  proton-proton collisions at $\sqrt{s} =  13\,\rm{ TeV}$ with a significance exceeding five standard deviations, the measured mass of the $B_c(\rm 2S)$ meson is $6871.0 \pm  1.2 \pm  0.8 \pm  0.8 \,\rm{ MeV}$ \cite{CMS-Bc-1902}. Furthermore, the CMS collaboration obtained the mass gaps $M_{B_c(\rm 2S)}-M_{B_c}=596.14\pm 1.2 \pm 0.8 \,\rm{MeV}$ and
$M_{B_c^*(\rm 2S)}-M_{B^*_c}=567.1\pm1.0\,\rm{MeV}$ \cite{CMS-Bc-1902}. Also recently,  the LHCb collaboration observed the $B_c^*(\rm 2S)$  in the $B_c^+\pi^+\pi^-$ invariant mass spectrum using proton-proton   collision data at $\sqrt{s} = 7$, $8$ and $13\,\rm{TeV}$, the measured mass is $6841.2 \pm 0.6  \pm 0.1  \pm 0.8 \,\rm{MeV}$ \cite{LHCb-Bc-1904}. The observations of the $B_c(\rm 2S)$ and $B_c^*(\rm 2S)$ states have an important implication in the mass spectrum of the $B_c$-like  tetraquark states.

In 2011, the Belle collaboration  observed  the $Z^\pm_b(10610)$ and $Z_b^\pm(10650)$ in the $\pi^{\pm}\Upsilon({\rm 1,2,3S})$  and $\pi^{\pm} h_b({\rm 1,2P})$ invariant  mass spectrum firstly, and determined the   favored   spin-parity  are $J^P=1^+$  \cite{Belle1105}, the  updated  masses and widths are $ M_{Z_b(10610)}=(10607.2\pm2.0)\,\rm{ MeV}$, $M_{Z_b(10650)}=(10652.2\pm1.5)\,\rm{MeV}$, $\Gamma_{Z_b(10610)}=(18.4\pm2.4) \,\rm{MeV}$ and
$\Gamma_{Z_b(10650)}=(11.5\pm2.2)\,\rm{ MeV}$, respectively \cite{Belle1110}.
The   $Z^\pm_b(10610)$ and $Z_b^\pm(10650)$ are excellent candidates for the hidden-bottom  tetraquark states \cite{Tetraquark-Zb,Tetraquark-Zb-QCDSR,WangHuang-2014-NPA}.

In 2013, the BESIII (also the Belle) collaboration  observed the $Z_c^{\pm}(3900)$ in the $\pi^\pm J/\psi$ invariant mass spectrum \cite{BES3900,Belle3900}.
   Furthermore, the BESIII collaboration  observed
the $Z^{\pm}_c(4025)$ near the $(D^{*} \bar{D}^{*})^{\pm}$ threshold \cite{BES1308},
and observed   the  $Z_c^{\pm}(4020)$   in the $\pi^\pm h_c$ invariant mass spectrum  \cite{BES1309}.  The $Z_c^\pm(3900)$ and $Z^\pm_c(4020/4025)$ are excellent candidates for the hidden-charm  tetraquark states \cite{Maiani1303,Tetraquark3900,WangHuangTao-3900,Nielsen3900}. For a recent comprehensive review of the $X$, $Y$, $Z$ particles both on the experimental and theoretical aspects, one can consult Ref.\cite{XYZ-review}.

If the $Z_c^\pm(3900)$, $Z_c^\pm(4020/4025)$, $Z^\pm_b(10610)$ and $Z_b^\pm(10650)$) are the diquark-antidiquark type charmonium-like and bottomonium-like tetraquark states, respectively, there should exist the diquark-antidiquark type $B_c$-like tetraquark states. It is interesting to  study this subject.
 At the charm sector, in 2014, the LHCb collaboration studied the $B^0\to\psi'\pi^-K^+$ decays    by performing a four-dimensional fit of the decay amplitude, and  provided the first independent confirmation of
the existence of the $Z_c^-(4430)$ state
and established its spin-parity to be $J^P=1^+$  \cite{LHCb-1404}.

We can assign  the  $Z_c(4430)$ to be the first radial excitation of the tetraquark candidate $Z_c(3900)$ according to the
analogous Okubo-Zweig-Iizuka supper-allowed decays,
\begin{eqnarray}
Z_c^\pm(3900)&\to&J/\psi\pi^\pm\, , \nonumber \\
Z_c^\pm(4430)&\to&\psi^\prime\pi^\pm\, ,
\end{eqnarray}
and analogous gaps $M_{Z_c(4430)}-M_{Z_c(3900)}=591\,\rm{MeV}$ and $M_{\psi^\prime}-M_{J/\psi}=589\,\rm{MeV}$  \cite{PDG,Z4430-1405,Nielsen-1401,Wang4430}.
In the QCD sum rules for the $Z_c(3900/4020)$ as the ground state axialvector tetraquark state, it is satisfactory to choose the continuum threshold parameters as $\sqrt{s_0}=M_{Z_c}+0.50/0.55\pm0.10\,\rm{GeV}$ \cite{WangZG-Z4600-Z4430}, as the energy gaps at the charm sector have the relation $M_{Z^\prime_c}-M_{Z_c}=M_{\psi^\prime}-M_{J/\psi}$. In Ref.\cite{WangZb10610-10650}, we assume    $M_{Z^\prime_b}-M_{Z_b}=M_{\Upsilon^\prime}-M_{\Upsilon}=0.55\,\rm{GeV}$ at the bottom sector, and choose the continuum threshold parameters as
 $\sqrt{s_0}=M_{Z_b}+0.55\pm0.10\,\rm{GeV}$ to calculate the mass spectrum of the hidden-bottom tetraquark states with the QCD sum rules. The precise value is $M_{\Upsilon^\prime}-M_{\Upsilon}=0.563\,\rm{GeV}$ from the Particle Data Group \cite{PDG}. For the vector $B_c$ mesons, $M_{B_c^*(\rm 2S)}-M_{B^*_c}=567 \,\rm{MeV}$ from the CMS collaboration \cite{CMS-Bc-1902}, and $M_{B_c^*(\rm 2S)}-M_{B^*_c}=566 \,\rm{MeV}$ from the LHCb collaboration \cite{LHCb-Bc-1904}.
Now we can take  the experimental data from the CMS and LHCb collaborations as input parameters and choose the continuum threshold parameters as   $\sqrt{s_0}=M_{Z_{\bar{b}c}}+0.55\pm0.10\,\rm{GeV}$ to study
the diquark-antidiquark type $B_c$-like tetraquark states with the QCD sum rules.

The $Z_b(10610)$, $Z_b(10650)$, $Z_c(3900)$ and $Z_c(4020)$ are observed in the analogous decays to the final states  $\pi^{\pm} h_b({\rm 1,2P})$, $\pi^{\pm}\Upsilon({\rm 1,2,3S})$, $\pi^\pm J/\psi$, $\pi^\pm h_c$, we expect that their  bottom-charm cousins $Z_{\bar{b}c}$ can be observed in the $B_c^* \pi^\pm$ mass spectrum.
Although in the decay $B_c^* \to B_c \gamma$,  the soft photon $\gamma$ is difficult to detect so as to reconstruct the $B_c^*$ state, the partial decay width $\Gamma(B_c^* \to B_c \gamma)$ is about $100\,\rm{eV}$ from the QCD sum rules \cite{WangBcVdecay}.  The   mass  splitting $M_{B_c^*}-M_{B_c}$ from the nonrelativistic renormalization group is about $46 \pm 15 {}^{+13}_{-0.11}  \,\rm{MeV}$ \cite{BcV-renorm}, which can be taken into account in the uncertainty analysis.

In previous works, we observed that the calculations based on the QCD sum rules  support  assigning the $Z_b(10610)$, $Z_b(10650)$, $Z_c(3900)$ and $Z_c(4020)$ to be the diquark-antidiquark type  axialvector tetraquark states, as there are more than one axialvector tetraquark candidates  for each of those $Z_b$ or $Z_c$ states \cite{WangHuang-2014-NPA,WangHuangTao-3900,WangZG-Z4600-Z4430,WangZb10610-10650,Wang-tetra-formula}.  If the dominant Fock components of those $Z_b$ or $Z_c$ states  are really diquark-antidiquark type tetraquark states, there should exist corresponding $B_c$-like tetraquark states. In this article, we study the axialvector $B_c$-like tetraquark states with the QCD sum rules. The observation of the $B_c$-like tetraquark states can shed light on the nature of  those $Z_b$ and $Z_c$ states, and plays
an important role in establishing  the tetraquark states.

The rest of the article is arranged as follows: in Sect.2, we obtain the QCD sum rules for
the masses and pole residues of the  axialvector $B_c$-like tetraquark states;
in Sect.3, we present the numerical results and detailed discussions; in Sect.4, we give a short conclusion.

\section{QCD sum rules for  the axialvector $B_c$-like tetraquark states}

Firstly, we write down  the two-point correlation functions  $\Pi_{\mu\nu}(p)$  in the QCD sum rules,
\begin{eqnarray}
\Pi_{\mu\nu}(p)&=&i\int d^4x e^{ip \cdot x} \langle0|T\Big\{J_\mu(x)J_{\nu}^{\dagger}(0)\Big\}|0\rangle \, ,
\end{eqnarray}
where $J_\mu(x)=J^{++,1}_{\pm,\mu}(x)$, $J^{+,1}_{\pm,\mu}(x)$, $J^{0,1}_{\pm,\mu}(x)$, $J^{+,0}_{\pm,\mu}(x)$,
\begin{eqnarray}
J^{++,1}_{\pm,\mu}(x)&=&\frac{\varepsilon^{ijk}\varepsilon^{imn}}{\sqrt{2}}\Big[u^{Tj}(x)C\gamma_5c^k(x) \bar{d}^m(x)\gamma_\mu C \bar{b}^{Tn}(x)\pm u^{Tj}(x)C\gamma_\mu c^k(x)\bar{d}^m(x)\gamma_5C \bar{b}^{Tn}(x) \Big] \, ,\nonumber\\
J^{+,1}_{\pm,\mu}(x)&=&\frac{\varepsilon^{ijk}\varepsilon^{imn}}{2}\Big[\Big(u^{Tj}(x)C\gamma_5c^k(x) \bar{u}^m(x)\gamma_\mu C \bar{b}^{Tn}(x) -d^{Tj}(x)C\gamma_5c^k(x) \bar{d}^m(x)\gamma_\mu C \bar{b}^{Tn}(x)\Big)  \nonumber\\
&&\pm \Big( u^{Tj}(x)C\gamma_\mu c^k(x)\bar{u}^m(x)\gamma_5C \bar{b}^{Tn}(x)-d^{Tj}(x)C\gamma_\mu c^k(x)\bar{d}^m(x)\gamma_5C \bar{b}^{Tn}(x)\Big)\Big]\, ,\nonumber\\
J^{0,1}_{\pm,\mu}(x)&=&\frac{\varepsilon^{ijk}\varepsilon^{imn}}{\sqrt{2}}\Big[d^{Tj}(x)C\gamma_5c^k(x) \bar{u}^m(x)\gamma_\mu C \bar{b}^{Tn}(x)\pm d^{Tj}(x)C\gamma_\mu c^k(x)\bar{u}^m(x)\gamma_5C \bar{b}^{Tn}(x) \Big] \, ,\nonumber\\
J^{+,0}_{\pm,\mu}(x)&=&\frac{\varepsilon^{ijk}\varepsilon^{imn}}{2}\Big[\Big(u^{Tj}(x)C\gamma_5c^k(x) \bar{u}^m(x)\gamma_\mu C \bar{b}^{Tn}(x) +d^{Tj}(x)C\gamma_5c^k(x) \bar{d}^m(x)\gamma_\mu C \bar{b}^{Tn}(x)\Big)  \nonumber\\
&&\pm \Big( u^{Tj}(x)C\gamma_\mu c^k(x)\bar{u}^m(x)\gamma_5C \bar{b}^{Tn}(x)+d^{Tj}(x)C\gamma_\mu c^k(x)\bar{d}^m(x)\gamma_5C \bar{b}^{Tn}(x)\Big)\Big]\, ,
\end{eqnarray}
  the $i$, $j$, $k$, $m$, $n$ are  color indexes, the $C$ is the charge conjugation matrix, the subscripts $\pm$ denote the positive and negative charge conjugation, respectively, the superscripts $++$, $+$, $0$ before the  comma  denote the electric charge, the superscripts $1$ and $0$ after the comma denote the isospin.  In the isospin limit, the currents $J_{+,\mu}(x)$ (or $J_{-,\mu}(x)$) couple to the diquark-antidiquark type axialvector $B_c$-like  tetraquark states with degenerate  masses. In the present work, we choose $J_\mu(x)=J^{+,0}_{\pm,\mu}(x)$ for simplicity.

At the hadron side, we   separate  the contributions of the ground state axialvector $B_c$-like tetraquark states,
\begin{eqnarray}
\Pi_{\mu\nu}(p)&=&\frac{\lambda_{Z}^2}{M_{Z}^2-p^2}\left( -g_{\mu\nu}+\frac{p_{\mu}p_{\nu}}{p^2}\right) +\cdots \nonumber\\
&=&\Pi(p^2)\left( -g_{\mu\nu}+\frac{p_{\mu}p_{\nu}}{p^2}\right)+\cdots \, ,
\end{eqnarray}
where the pole residues $\lambda_{Z}$ are defined by
\begin{eqnarray}
 \langle 0|J_\mu(0)|Z_{\bar{b}c}(p)\rangle &=&\lambda_{Z}\, \varepsilon_\mu\, ,
\end{eqnarray}
the  $\varepsilon_{\mu}$ is the polarization vector.

At the QCD side, we take into account the vacuum condensates  $ \langle\bar{q}q\rangle$, $\langle\frac{\alpha_{s}GG}{\pi}\rangle$,
$\langle\bar{q}g_s\sigma Gq\rangle$, $\langle\bar{q}q\rangle^2$, $\langle\bar{q}q\rangle\langle\frac{\alpha_{s}GG}{\pi}\rangle$,
$\langle\bar{q}q\rangle\langle\bar{q}g_s\sigma Gq\rangle$, $\langle\bar{q}g_s\sigma Gq\rangle^2$ and  $\langle\bar{q}q\rangle^2\langle\frac{\alpha_{s}GG}{\pi}\rangle$          in the operator product expansion  according to routines  in Refs.\cite{WangHuangTao-3900,Wang-tetra-formula,WangZG-molecule}, obtain the QCD spectral densities $\rho(s)$ through dispersion relation. Then we match  the hadron side with the QCD  side of the components $\Pi(p^2)$   and perform Borel transform   to obtain  the  QCD sum rules:
\begin{eqnarray}\label{QCDSR}
\lambda^2_{Z}\, \exp\left(-\frac{M^2_{Z}}{T^2}\right)= \int_{(m_b+m_c)^2}^{s_0} ds\, \rho(s) \, \exp\left(-\frac{s}{T^2}\right) \, ,
\end{eqnarray}
where the $s_0$ is  the continuum threshold parameter, the $T^2$ is the Borel parameter.
 The lengthy  expressions of the QCD spectral densities $\rho(s)$ are neglected for simplicity. In the QCD sum rules for the tetraquark states consist of two heavy quarks and two light quarks, we have to carry out the operator product expansion up to the vacuum condensates of dimension $10$. Because there are four quark lines in the correlation functions, if each
heavy quark line emits a gluon, each light quark line contributes a $\bar{q}q$ pair, we obtain a operator $\bar{q}q\bar{q}qG_{\mu\nu}G_{\alpha\beta}$ in the fixed point gauge, the operator is of dimension $10$, see Fig.1.
\begin{figure}
\centering
\includegraphics[totalheight=4cm,width=6cm]{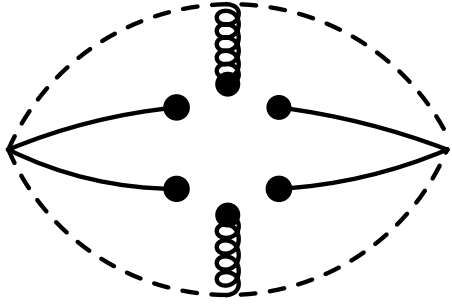}
  \caption{ A typical Feynman diagram contributes to the  vacuum condensates of dimension $10$, where the solid and dashed lines denote the light and heavy quarks, respectively.  }
\end{figure}

We derive Eq.\eqref{QCDSR} with respect to  $\tau=\frac{1}{T^2}$,  and obtain the QCD sum rules for  the masses,
 \begin{eqnarray}
 M^2_{Z}&=& -\frac{\frac{d}{d \tau} \int_{(m_b+m_c)^2}^{s_0} ds\rho(s)\exp\left(-\tau s \right)}{\int_{(m_b+m_c)^2}^{s_0} ds \rho(s)\exp\left(-\tau s\right)}\, .
\end{eqnarray}

\section{Numerical results and discussions}

We choose   the popular or conventional  values of the vacuum condensates $\langle
\bar{q}q \rangle=-(0.24\pm 0.01\, \rm{GeV})^3$,   $\langle
\bar{q}g_s\sigma G q \rangle=m_0^2\langle \bar{q}q \rangle$,
$m_0^2=(0.8 \pm 0.1)\,\rm{GeV}^2$,  $\langle \frac{\alpha_s
GG}{\pi}\rangle=(0.33\,\rm{GeV})^4 $    at the energy scale  $\mu=1\, \rm{GeV}$
\cite{SVZ79,Reinders85,Colangelo-Review}. As far as the heavy quark masses are concerned, the values of the heavy quark pole masses are $\hat{m}_c=1.67 \pm 0.07 \,\rm{GeV }$ and $\hat{m}_b=4.78\pm0.06 \,\rm{GeV }$ from the Particle Data Group \cite{PDG}, which lead to the relations $2\hat{m}_c>M_{\eta_c}=2.9839\,\rm{GeV}$ and
  $2\hat{m}_b>M_{\eta_b}=9.399\,\rm{GeV}$. So  we cannot choose the pole masses in the QCD sum rules.
Now we only have one option to choose the $\overline{MS}$ masses $m_{c}(m_c)=(1.275\pm0.025)\,\rm{GeV}$ and $m_{b}(m_b)=(4.18\pm0.03)\,\rm{GeV}$ from the Particle Data Group \cite{PDG}. Furthermore, we neglect the small $u$ and $d$ quark masses.
Then we take into account the energy-scale dependence of  the input parameters from the renormalization group equation,
\begin{eqnarray}
\langle\bar{q}q \rangle(\mu)&=&\langle\bar{q}q \rangle({\rm 1GeV})\left[\frac{\alpha_{s}({\rm 1GeV})}{\alpha_{s}(\mu)}\right]^{\frac{12}{33-2n_f}}\, , \nonumber\\
 \langle\bar{q}g_s \sigma Gq \rangle(\mu)&=&\langle\bar{q}g_s \sigma Gq \rangle({\rm 1GeV})\left[\frac{\alpha_{s}({\rm 1GeV})}{\alpha_{s}(\mu)}\right]^{\frac{2}{33-2n_f}}\, , \nonumber\\
 m_c(\mu)&=&m_c(m_c)\left[\frac{\alpha_{s}(\mu)}{\alpha_{s}(m_c)}\right]^{\frac{12}{33-2n_f}} \, ,\nonumber\\
  m_b(\mu)&=&m_b(m_b)\left[\frac{\alpha_{s}(\mu)}{\alpha_{s}(m_b)}\right]^{\frac{12}{33-2n_f}} \, ,\nonumber\\
\alpha_s(\mu)&=&\frac{1}{b_0t}\left[1-\frac{b_1}{b_0^2}\frac{\log t}{t} +\frac{b_1^2(\log^2{t}-\log{t}-1)+b_0b_2}{b_0^4t^2}\right]\, ,
\end{eqnarray}
   where $t=\log \frac{\mu^2}{\Lambda^2}$, $b_0=\frac{33-2n_f}{12\pi}$, $b_1=\frac{153-19n_f}{24\pi^2}$, $b_2=\frac{2857-\frac{5033}{9}n_f+\frac{325}{27}n_f^2}{128\pi^3}$,  $\Lambda=210\,\rm{MeV}$, $292\,\rm{MeV}$  and  $332\,\rm{MeV}$ for the flavors  $n_f=5$, $4$ and $3$, respectively  \cite{PDG,Narison-mix}. In the present work, we study the $B_c$-like tetraquark states, it is better to choose the flavor $n_f=5$. Now we begin to discuss how to choose the ideal energy scales of the QCD spectral densities.

In the QCD sum rules for  the conventional (two-quark) mesons and (three-quark) baryons,
the QCD spectral densities have the form  $\rho(s)\sim s^n$ with $n\leq 1$ for the mesons and $n\leq 2$ for the baryons in the zero quark mass limit. The convergent behavior of the operator product expansion is very good. While in the QCD sum rules for the tetraquark states, irrespective of the diquark-antidiquark type compact tetraquark states and meson-meson type molecular states, the QCD spectral densities have the form  $\rho(s)\sim s^n$ with $n\leq 4$ in the zero quark mass limit, the convergent behavior of the operator product expansion is bad, we have to choose large Borel parameter to suppress the contributions of the higher dimensional vacuum condensates.

At the hadron side, if we choose  the "single-pole + continuum states" model to represent the hadron spectral densities to study the ground states, we have to calculate the contributions of the vacuum condensates $D(n)$ of dimension $n$ with the formula,
\begin{eqnarray}
D(n)&=&\frac{\int_{\Delta^{2}}^{s_{0}}ds\rho_{n}(s)\exp\left(-\frac{s}{T^{2}}\right)}
{\int_{\Delta^{2}}^{s_{0}}ds\rho\left(s\right)\exp\left(-\frac{s}{T^{2}}\right)}\, ,
\end{eqnarray}
rather than with the formula,
\begin{eqnarray}
D(n)&=&\frac{\int_{\Delta^{2}}^{\infty}ds\rho_{n}(s)\exp\left(-\frac{s}{T^{2}}\right)}
{\int_{\Delta^{2}}^{\infty}ds\rho\left(s\right)\exp\left(-\frac{s}{T^{2}}\right)}\, , \label{Dninfty}
\end{eqnarray}
where the $\Delta^2$ denotes the lower thresholds. For the hidden-charm or hidden-bottom tetraquark states and molecular states, the lower thresholds $\Delta^2=4m_Q^2$,   the QCD spectral densities $\rho(s)$ also depend on the heavy quark mass $m_Q$ heavily,  small variation of the $m_Q$ can lead to rather different result after carrying out the integral over  $ds$,  if the upper threshold $s_0$  is chosen.   In the following, we will take the QCD sum rules for the hidden-charm (hidden-bottom) tetraquark states as an example to illustrate how to choose the ideal energy scales.

 The  heavy quark ($\overline{MS}$) mass $m_Q(\mu)$ depends on the energy scale in QCD, it is an energy scale dependent quantity, irrespective of whether or not we calculate the perturbative $\mathcal{O}(\alpha_s,\alpha_s^2,\alpha_s^3,\cdots)$ corrections in a particular QCD sum rule, as the QCD sum rule in itself  is a QCD approach.

We can write the correlation functions $\Pi(p^2)$ at the QCD side as
\begin{eqnarray}
\Pi(p^2)&=&\int_{4m^2_Q(\mu)}^{s_0} ds \frac{\rho(s,\mu)}{s-p^2}+\int_{s_0}^\infty ds \frac{\rho(s,\mu)}{s-p^2} \, ,
\end{eqnarray}
which are  scale independent quantities,
\begin{eqnarray}
\frac{d}{d\mu}\Pi(p^2)&=&0\, ,
\end{eqnarray}
we can carry out the operator product expansion  at any energy scales at which perturbative calculations are feasible.
In practical calculations, we cannot calculate the perturbative corrections up to arbitrary orders, even the next-to-leading order, and have to make truncations in one way or the other. Furthermore, we  have to factorize the  higher dimensional vacuum condensates  into lower dimensional ones paying  the price of modifying the energy scale dependence, as our knowledge on the higher dimensional vacuum condensates are scarce. The truncation $s_0$ for the continuum contributions   makes the situation even bad, as the correlation between the threshold $4m^2_Q(\mu)$ and continuum threshold $s_0$ is unknown. So we cannot obtain energy scale independent QCD sum rules,
\begin{eqnarray}
\frac{d}{d\mu}\int_{4m^2_Q(\mu)}^{s_0} ds \frac{\rho(s,\mu)}{s-p^2}&\rightarrow& \frac{d}{d\mu}\int_{4m^2_Q(\mu)}^{s_0} ds \frac{\rho(s,\mu)}{T^2}\exp\left( -\frac{s}{T^2}\right)\neq 0 \, .
\end{eqnarray}
We cannot extract the hadron (or tetraquark) masses from energy scale independent QCD sum rules, as the QCD spectral densities $\rho(s,\mu)$ depend on the energy scales, the thresholds
$4m_Q^2(\mu)$ also depend on the energy scales.
Even in the QCD sum rules for the (conventional) pseudoscalar $D$ and $B$ mesons, where the perturbative $\mathcal{O}(\alpha_s^2)$ corrections to the perturbative terms are calculated \cite{QCDSR-3loop}, the perturbative $\mathcal{O}(\alpha_s)$ corrections to the quark condensate terms are also calculated \cite{QCDSR-condensare-1loop}, we still cannot obtain energy scale  independent QCD sum rules.

We can study  the hidden-bottom or hidden-charm tetraquark states with    a double-well potential model.
   In the heavy quark limit, the heavy  quark $Q$ serves as a static well potential,
which attracts  the light quark $q$ to form a diquark in the color antitriplet channel. While the heavy  antiquark $\overline{Q}$ serves as
another static well potential, which attracts  the light antiquark $\bar{q}$ to form a antidiquark in the color triplet channel.
We can introduce the effective heavy quark masses ${\mathbb{M}}_Q$ and the virtuality $V=\sqrt{M^2_{X/Y/Z}-(2{\mathbb{M}}_Q)^2}$ to
 describe  those  tetraquark states.

Now the QCD sum rules for the hidden-charm or hidden-bottom tetraquark states have three typical energy scales $\mu^2$, $T^2$, $V^2$,
 it is natural to choose the energy  scale \cite{WangHuang-2014-NPA,Wang-tetra-formula},
 \begin{eqnarray}
 \mu^2&=&V^2={\mathcal{O}}(T^2)\, ,
 \end{eqnarray}
then we obtain the formula $\mu=\sqrt{M^2_{X/Y/Z}-(2{\mathbb{M}}_Q)^2}$ to choose the ideal energy scales of the QCD spectral densities. At the ideal energy scales, we can enhance the pole contributions at the hadron side remarkably and improve the convergent behaviors of the operator product expansion at the QCD side remarkably. For the hidden-bottom and hidden-charm tetraquark states, we can obtain the  pole contributions as large as $(40-60)\%$, or even larger. Thus we can avoid to extract the tetraquark  masses at small pole contributions and obtain more reliable predictions. Otherwise, we have to resort to  the "multi-pole $+$ continuum states" model to approximate the
hadronic   spectral densities and postpone the continuum threshold parameters to very large values.

In the present case, there are a $c$-quark and a $b$ quark in the tetraquark states, we modify the energy scale formula to be
\begin{eqnarray}
\mu&=&\sqrt{M^2_{Z}-({\mathbb{M}}_b+{\mathbb{M}}_c)^2}\, ,
\end{eqnarray}
with the updated effective heavy quark masses ${\mathbb{M}}_b=5.17\,\rm{GeV}$ and ${\mathbb{M}}_c=1.82\,\rm{GeV}$ to determine the ideal energy scales of the QCD spectral densities \cite{WangZb10610-10650,WangEPJC1601}. We can rewrite the energy scale formula as $M_Z=\sqrt{\mu^2+({\mathbb{M}}_b+{\mathbb{M}}_c)^2}$, and choose the lowest feasible energy scale $\mu=1\,\rm{GeV}$ to make a crude estimation for the masses  of the axialvector $B_c$-like tetraquark states, $M_Z\geq 7.06\,\rm{GeV}$. The crude estimation is based on our previous works on the masses (and widths) of the $Z_c(3900)$, $Z_c(4020)$, $Z_c(4430)$, $Y(4600)$, $Z_b(10610)$, $Z_b(10650)$, etc as the diquark-antiquark type tetraquark states using the QCD sum rules \cite{WangHuang-2014-NPA,WangHuangTao-3900,WangZG-Z4600-Z4430,WangZb10610-10650,Wang-tetra-formula,WangEPJC1601,WangZhang-solid-1,WangZhang-solid-2}.
 In Ref.\cite{DEbert-tetra}, D. Ebert et al study the mass spectrum  of the diquark-antidiquark type tetraquark states with two heavy quarks
 in the relativistic quasipotential quark model, and obtain the axialvector tetraquark masses, about $7.20-7.24\,\rm{GeV}$, which are consistent with the present estimation.

Now we search for the ideal   Borel parameters $T^2$ and continuum threshold parameters $s_0$  to obey   the  four criteria:\\
$\bf 1.$ Pole dominance at the hadron side;\\
$\bf 2.$ Convergence of the operator product expansion at the QCD side;\\
$\bf 3.$ Appearance of the Borel platforms;\\
$\bf 4.$ Satisfying the energy scale formula,\\
 via  try and error. For the continuum threshold parameters $s_0$, we put  an additional constraint $\sqrt{s_0}=M_{Z_{\bar{b}c}}+0.55\pm0.10\,\rm{GeV}$ considering the experimental data from the CMS and LHCb collaborations  \cite{CMS-Bc-1902,LHCb-Bc-1904}.

Although the searching process is very long,  we obtain the Borel parameters (or Borel windows) $T^2$, continuum threshold parameters $s_0$, ideal energy scales of the QCD spectral densities,  pole contributions, and the contributions of the vacuum condensates of dimension $10$, which are shown explicitly in Table \ref{BorelP}.
The pole contributions are about $(43-63)\%$ in the Borel windows, the central values exceed $50\%$,  the pole dominance condition  can be satisfied.
On the other hand, the contributions of the vacuum condensates of dimension $10$  are less than $2\%$ in the Borel windows, the operator product expansion is well convergent.

\begin{figure}
\centering
\includegraphics[totalheight=6cm,width=7cm]{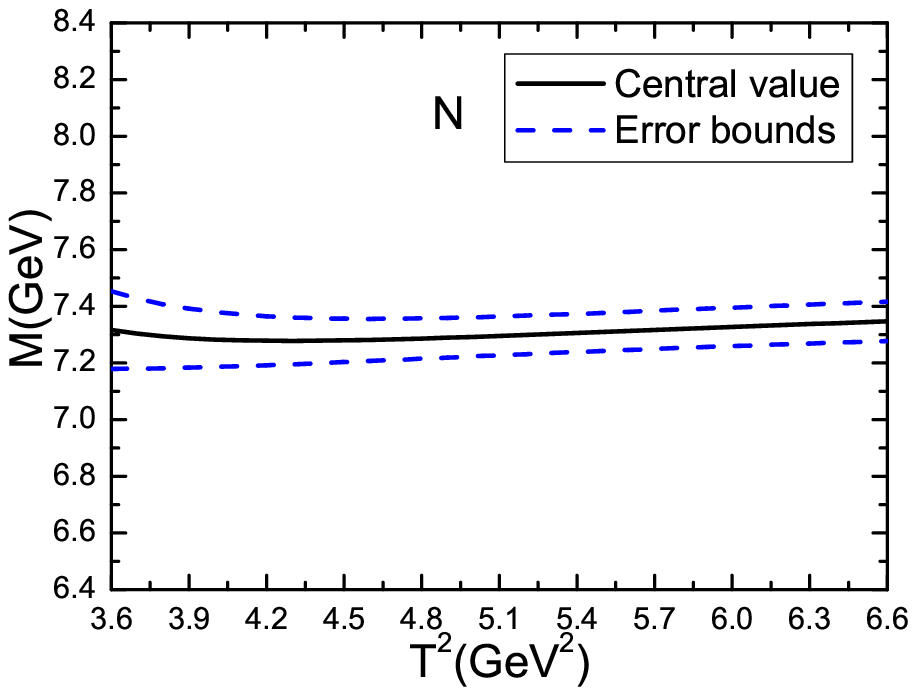}
\includegraphics[totalheight=6cm,width=7cm]{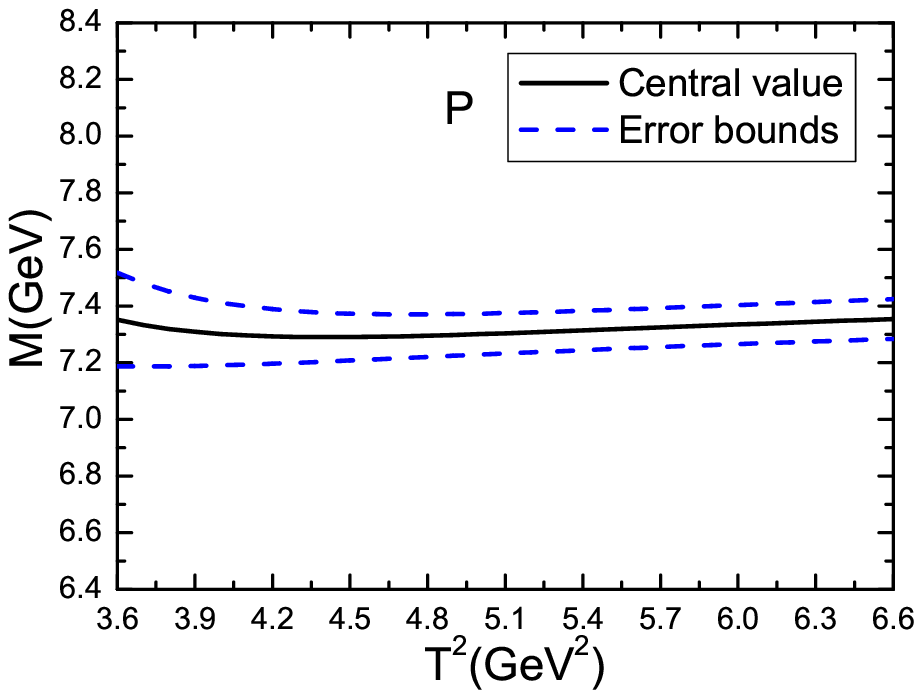}\
  \caption{ The masses  with variations of the  Borel parameters $T^2$ for  the axialvector $B_c$-like tetraquark states, where the $N$ and $P$ denote the negative and positive  charge conjugation, respectively. }\label{mass-1-fig}
\end{figure}

\begin{figure}
\centering
\includegraphics[totalheight=6cm,width=7cm]{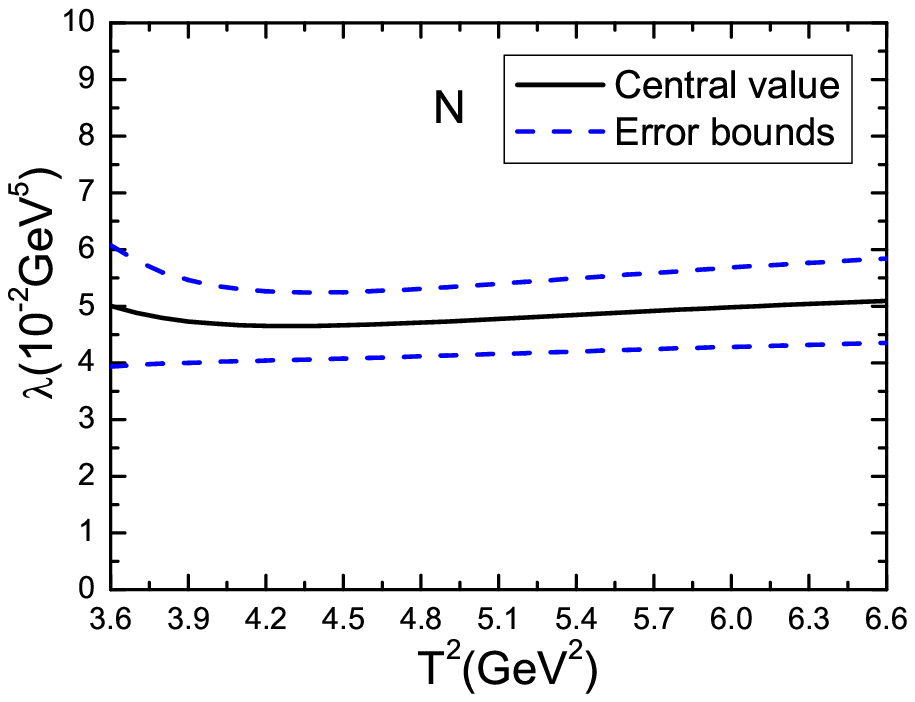}
\includegraphics[totalheight=6cm,width=7cm]{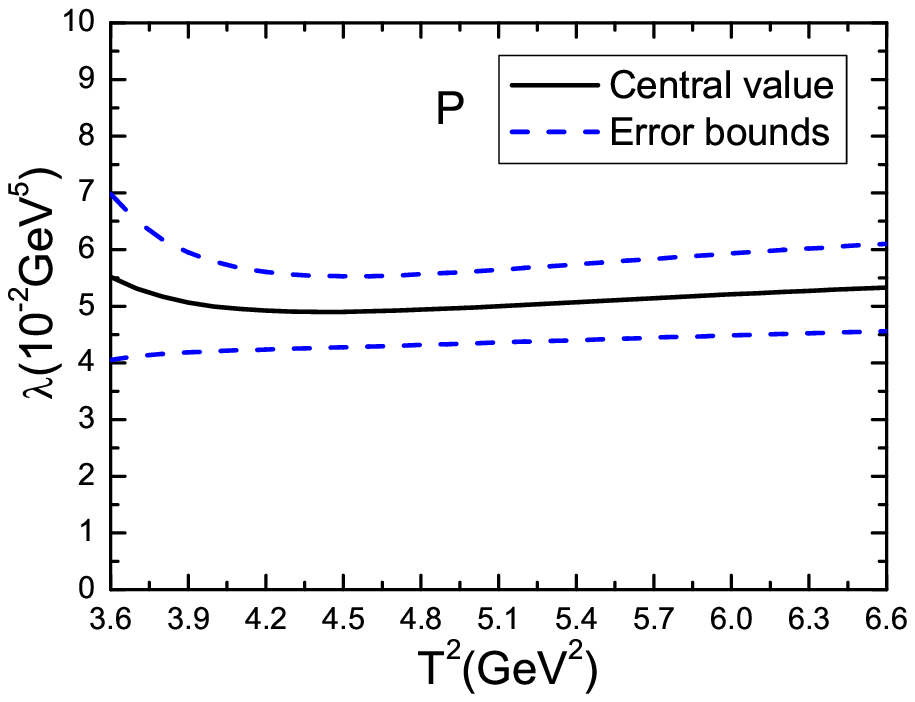}\
  \caption{ The pole residues   with variations of the  Borel parameters $T^2$ for  the axialvector $B_c$-like tetraquark states, where the $N$ and $P$ denote the negative and positive  charge conjugation, respectively. }\label{residue-1-fig}
\end{figure}

Then we take into account  the uncertainties of the input parameters and obtain the masses and pole residues of the  axialvector $B_c$-like  tetraquark states, which are shown explicitly in Table \ref{mass-Table} and in Figs.\ref{mass-1-fig}--\ref{residue-1-fig}. From  Tables \ref{BorelP}--\ref{mass-Table}, we can see that the energy scale formula $\mu=\sqrt{M^2_{Z}-({\mathbb{M}}_b+{\mathbb{M}}_c)^2}$ is well satisfied.
 In  Figs.\ref{mass-1-fig}--\ref{residue-1-fig}, we plot the masses and pole residues of the  axialvector $B_c$-like tetraquark states with variations of the Borel parameters at much larger ranges than the Borel widows, in the Borel windows, the Borel platforms appear.  Now the four criteria of the QCD sum rules are all satisfied.
and we expect to make reasonable predictions.

The $DB^*$ and $D^*B$ thresholds are $7192\,\rm{MeV}$ and $7288\,\rm{MeV}$, respectively,  the predicted masses  $M_{Z_{\bar{b}c}(1^{+-})}=7.30\pm0.08\,\rm{GeV}$ and $M_{Z_{\bar{b}c}(1^{++})}=7.31\pm0.08\,\rm{GeV}$, which lie above the $DB^*$ threshold, the fall-apart decays  $Z_{\bar{b}c}\to DB^*$ can take place kinematically, while the corresponding decays $Z_{\bar{b}c}\to D^*B$ can take place marginally.
Although the $Z_b(10610)$ and $Z_b(10650)$ are observed in the decays $Z^+_b(10610/10650) \to \pi^+\Upsilon({\rm 1,2,3S})$ and $\pi^+ h_b({\rm 1,2P})$, the dominant
decay modes are $Z_b^+(10610)\to B^+\bar{B}^{*0}+\bar{B}^0B^{*+}$ and $Z^+_b(10650)\to B^{*+}\bar{B}^{*0}$ \cite{Exp-Zb-decays}. At the charm sector, the dominant decay modes of the corresponding hidden-charm tetraquark candidates are $Z_c^\pm(3900/3885)\to (D\bar{D}^*)^\pm$ \cite{Exp-Zc3900-decay}, and $X(3872)\to D^{*0}\bar{D}^0+D^{0}\bar{D}^{*0}$ \cite{YuanCZ-decay}. We can search for the $Z_{\bar{b}c}$ states in the Okubo-Zweig-Iizuka supper-allowed decays $Z_{\bar{b}c}\to B_c^*\pi$, $B_c\rho$, $B_c^*\rho$, $B_c\pi$, $DB^*$ and $D^*B$, and compare  the present predictions with the experimental data  at the LHCb, Belle II,  CEPC, FCC, ILC
 in the future, which maybe  shed light on the nature of the exotic $X$, $Y$, $Z$ particles.

If we take the heavy quark limit, and choose the same hadronic coupling constants in the  strong decays
$Z_c(3900)\to J/\psi \pi$, $\eta_c\rho$, $D\bar{D}^*$, $D^*\bar{D}$ to study the analogous    decays $Z_{\bar{b}c}(1^{+-})\to B_c^*\pi$, $B_c\rho$, $DB^*$ and $D^*B$ \cite{WangZhang-solid-1}, we can obtain a decay width $\Gamma_{Z_{\bar{b}c}(1^{+-})}\approx 20\,\rm{MeV}$,  which is  much smaller than the width $ 370^{+70}_{-70}{}^{+70}_{-132}\,\rm{MeV}$ of the $Z_c(4200)$ \cite{PDG,Wang-Octet}.
 In Ref.\cite{Wang-Octet}, we observe that  the finite width  effect can be absorbed   into the pole residue $\lambda_{Z_c(4200)}$ safely, the contributions of the scattering states  cannot  affect  the mass $M_{Z_c(4200)}$ significantly,  the zero width and single pole approximation  works, even if $\Gamma_{Z_{\bar{b}c}(1^{++})}>\Gamma_{Z_{\bar{b}c}(1^{+-})}$.

\begin{table}
\begin{center}
\begin{tabular}{|c|c|c|c|c|c|c|c|c|}\hline\hline
 $Z_{\bar{b}c}$                                       & $J^{PC}$ & $T^2 (\rm{GeV}^2)$ & $\sqrt{s_0}(\rm GeV) $      &$\mu(\rm{GeV})$   &pole         &$D(10)$ \\ \hline

$[qc]_S[\overline{qb}]_{A}-[qc]_{A}[\overline{qb}]_S$ & $1^{+-}$ & $5.0-5.6$          & $7.85\pm0.10$               &$2.10$            &$(43-63)\%$  &$<2\%$    \\

$[qc]_S[\overline{qb}]_{A}+[qc]_{A}[\overline{qb}]_S$ & $1^{++}$ & $5.0-5.6$          & $7.86\pm0.10$               &$2.15$            &$(43-63)\%$  &$\leq1\%$    \\
\hline\hline
\end{tabular}
\end{center}
\caption{ The Borel windows, continuum threshold parameters, energy scales of the QCD spectral densities,  pole contributions, and the contributions of the vacuum condensates of dimension $10$  for the ground state $B_c$-like tetraquark states. }\label{BorelP}
\end{table}

\begin{table}
\begin{center}
\begin{tabular}{|c|c|c|c|c|c|c|c|c|}\hline\hline
 $Z_{\bar{b}c}$                                                         & $J^{PC}$  & $M_Z (\rm{GeV})$   & $\lambda_Z (\rm{GeV}^5) $             \\ \hline

$[qc]_S[\overline{qb}]_{A}-[qc]_{A}[\overline{qb}]_S$                   & $1^{+-}$  & $7.30\pm0.08$      & $(4.82\pm0.71)\times 10^{-2}$        \\

$[qc]_S[\overline{qb}]_{A}+[qc]_{A}[\overline{qb}]_S$                   & $1^{++}$  & $7.31\pm0.08$      & $(5.05\pm0.73)\times 10^{-2}$        \\ \hline\hline
\end{tabular}
\end{center}
\caption{ The masses and pole residues of the ground state $B_c$-like tetraquark states. }\label{mass-Table}
\end{table}

The calculations based on the simple chromomagnetic interactions indicate that the lowest axialvector $B_c$-like tetraquark state has a mass about $6.928\,\rm{GeV}$ \cite{LYR-CS-Bc-tetra}, which is below the masses $7.11-7.15\,\rm{GeV}$  for the conventional  $2\rm{P}$ $B_c$ states with  $J^P=1^+$
in the potential quark models  \cite{GI,EFG,GJ,ZVR,GKLT,EQ,LuQF-potential}. The masses of the  conventional  $3\rm{P}$ $B_c$ states with  $J^P=1^+$
from the potential quark models are about  $7.45\,\rm{GeV}$ \cite{ZVR,LuQF-potential}.
The masses $M_{Z_{\bar{b}c}(1^{+-})}=7.30\pm0.08\,\rm{GeV}$ and $M_{Z_{\bar{b}c}(1^{++})}=7.31\pm0.08\,\rm{GeV}$ lie between the conventional $2\rm{P}$ and $3\rm{P}$ $B_c$ mesons with $J^P=1^+$, which is a typical feature of the diquark-antidiquark type $B_c$-like tetraquark states.

In Refs.\cite{Chen-W-Bc,Azizi-Bc-tetra}, the calculations based on the QCD sum rules with a different input parameter scheme indicate that the diquark-antidiquark ($6_c\otimes \bar{6}_c$) type axialvector $B_c$-like tetraquark states have  the masses about $7.10\,\rm{GeV}$ or $7.06\,\rm{GeV}$, which  differ from the present work completely, as we study the $\bar{3}_c\otimes 3_c $ type $B_c$-like tetraquark states.

\section{Conclusion}
In this article,  we construct the diquark-antidiquark type current operators to study the  masses and pole residues of the axialvector $B_c$-like tetraquark states  with
the QCD sum rules  by carrying out the operator product expansion up to vacuum condensates of dimension $10$. In calculations, we take the energy scale formula $\mu=\sqrt{M^2_{Z}-({\mathbb{M}}_b+{\mathbb{M}}_c)^2}$ as a powerful constraint to determine the ideal energy scales of the QCD spectral densities and give detailed discussions to illustrate  why we take the energy scale formula to improve the QCD sum rules for the doubly heavy tetraquark states. The present predictions depend heavily on the assignments  of the  $Z_c(3900)$, $Z_c(4020)$, $Z_c(4430)$, $Y(4600)$, $Z_b(10610)$, $Z_b(10650)$, etc as the diquark-antidiquark type tetraquark states, and the experimental data about the $B_c^*(\rm 2S)$ state from the CMS and LHCb collaborations.
The predicted masses $M_{Z_{\bar{b}c}(1^{+-})}=7.30\pm0.08\,\rm{GeV}$ and $M_{Z_{\bar{b}c}(1^{++})}=7.31\pm0.08\,\rm{GeV}$ lie between the conventional $2\rm{P}$ and $3\rm{P}$ $B_c$ mesons with $J^P=1^+$, which is a typical feature of the diquark-antidiquark type $B_c$-like tetraquark states.
We can compare   the present  predictions  with  the experimental data in the future at the LHCb, Belle II,  CEPC, FCC, ILC to diagnose the nature of the $X$, $Y$, $Z$ states.

\section*{Acknowledgements}
This  work is supported by National Natural Science Foundation, Grant Number  11775079.

\end{document}